\documentclass[]{emulateapj}
\usepackage{epsf}

\def\mbf#1{\mbox{\boldmath ${#1}$}}
\def\Alfven{Alfv\'{e}n~}
\def\Alfvenic{Alfv\'{e}nic~}

\slugcomment{Submitted to ApJ}

\shorttitle{Fast-mode Turbulence}
\shortauthors{Suzuki, Lazarian, \& Beresnyak}

\begin{document}

\title{Cascading of Fast-Mode Balanced and Imbalanced Turbulence}

\author{T. K. Suzuki$^1$, A. Lazarian$^2$, A. Beresnyak$^2$}
\altaffiltext{1}{School of Arts and Sciences, University of Tokyo, 
3-8-1, Komaba, Meguro, Tokyo, 153-8902, Japan}
\altaffiltext{2}{Department of Astronomy, University of Wisconsin,
475 N. Charter St., Madison, WI 53706
}

\email{stakeru@ea.c.u-tokyo.ac.jp}
\email{lazarian, andrey@astro.wisc.edu}

\begin{abstract}
We study the cascading of fast MHD modes in magnetically dominated plasma 
by performing one-dimensional (1D) dynamical simulations.
We find that the cascading becomes more efficient as an angle
between wave vector and underlying magnetic field increases and
fast mode becomes more compressive. We also study imbalanced
turbulence, in which wave packets propagating in one-direction have
more power than those in the opposite direction.  Unlike imbalanced
\Alfvenic turbulence, the imbalanced fast mode turbulence shows faster
cascading as the degree of imbalance increases.
We find that the spectral index of the velocity and magnetic field, which 
are carried by the fast-mode turbulence, quickly reaches stationary value 
of $-2$. 
Thus we conclude that the dissipation of fast mode, at least in
1D case, happens not due to weak or strong turbulent cascading, but
mostly due to nonlinear steepening.
The density fluctuations, which are carried by slow-mode perturbation 
in the larger scale and entropy-mode perturbation in the smaller scale, 
depend on initial driving spectrum 
and a ratio of specific heat. 

\end{abstract}

\section{Introduction}

Magnetohydrodynamical (MHD) turbulence is ubiquitous in astrophysics.
All interstellar medium (ISM) phases are well magnetized with Larmor
radius of thermal proton much smaller than outer scale.
Molecular clouds are no exception, with compressible, often
high-Mach number, magnetic
turbulence determining most of their properties (see
Elmegreen \& Falgarone 1996, Stutzki 2001).
Star formation (see McKee \& Tan 2002; Elmegreen 2002; 
Pudritz 2002; Ballesteros-Paredes et al.2006), 
cloud chemistry (see Falgarone 1999 and references therein), shattering and
coagulation of dust (see Lazarian \& Yan 2002 and references therein) 
are examples
of processes for which knowledge of turbulence is absolutely
essential.    
It is also believed that MHD waves and turbulence are important
for the acceleration of the solar wind (e.g. Tu \& Marsch 1995) as well as 
winds from stars which possess surface convective layers (e.g. Tsuji 1988).

As MHD turbulence is an extremely complex process, all theoretical
constructions should be tested thoroughly. Until very recently matching
of observations with theoretical constructions used to be the only way
of testing. Indeed, it is very dangerous to do MHD turbulence testing
without fully 3D MHD simulations with a distinct inertial range.
Theoretical advances related to the anisotropy of MHD turbulence
and its scalings (see Shebalin et al. 1983, Higdon 1984, Montgomery,
Brawn \& Matthaeus 1987, Shebalin \& Montgomery 1988, Zank \& Matthaeus
1992) were mostly done in relation with the observations of fluctuations
in solar wind. Computers allowed an important alternative
way of dealing with the problem. While still presenting a limited
inertial range, they allow to control the input parameters making it
easier to test theoretical ideas.

It is well known that linear MHD perturbations can be decomposed into
Alfv\'{e}nic, slow, fast, and entropy modes with well-defined dispersion
relations (see Alfv\'{e}n \& F\"{a}lthammar 1963). The separation into
Alfv\'en and pseudo-Alfv\'en modes, which are the incompressible limit
of Alfv\'en and slow MHD modes, is an essential element of the 
Goldreich-Sridhar (1995, henceforth GS95) model of turbulence.  
There the arguments were
provided in favor of Alfv\'en modes developing a cascade on their own,
while the pseudo-Alfv\'en modes being passively advected by the
cascade. The drain of Alfv\'enic mode energy to pseudo-Alfv\'en modes
was shown to be marginal.

The separation of 
MHD perturbations into modes in compressible media was discussed
further in Lithwick \& Goldreich (2001) and 
Cho \& Lazarian (2002, 2003
henceforth CL02, CL03, respectively). Even though MHD turbulence is 
a highly non-linear phenomenon,
the modes does not constitute an entangled inseparable mess.
 
The actual decomposition of MHD turbulence into 
each mode was a challenge that was addressed in CL02, CL03. 
They studied a particular realization of turbulence
with mean magnetic field comparable to the fluctuating magnetic field.
This setting is, on one hand, rather typical of the most MHD flows in Galaxy
and, on the other hand, allows them to use a statistical procedure of
decomposition in the Fourier space, where the bases of the Alfv\'en, slow,
fast, and entropy perturbations were defined. The entirely different 
way to decompose modes in high or low-$\beta$ cases in real space
was shown in CL03 to correspond well to this procedure\footnote{
In CL03, an isothermal equation of state is used so that entropy-mode 
fluctuation is prohibited.}.

 In particular, calculations in CL03 demonstrated that fast modes are
 marginally affected by Alfv\'{e}n modes and follow acoustic cascade
 in both high and low $\beta$ medium.  In Yan \& Lazarian (2002; 2004; 
henceforth YL04) the fast modes were identified as the
 major agent in scattering of cosmic rays. Similar results in relation
 to stochastic acceleration of cosmic rays were reached in Cho \&
 Lazarian (2006). Interestingly enough, the dominance of fast modes
 for both acceleration and scattering is present in spite of the
 dissipative character of fast modes. The reason for that is
 isotropy\footnote{ Recent calculations of fast modes interacting with
   Alfv\'{e}n modes by Chandran (2005) indicate some anisotropy of fast
   modes. In particular, he shows that contours of isocorrelation for
   fast modes get oblate with shorter direction along the magnetic
   field. However, the calculations are performed assuming that Alfv\'{e}n
   modes are in the weak turbulence regime.  This regime has a limited
   inertial range. Moreover, anisotropic damping discussed in YL04 is
   probably a more strong source of anisotropy. Anyhow, the above
   conclusions about the dominance of fast modes are not altered by
   the anisotropies that corresponds to squashing contours of
   isocorrelations in the direction of magnetic field.} of this mode
 reported in CL03. The Alfv\'{e}n modes, that are still the default for
 scattering and acceleration for many researchers, are inefficient due
 to the elongation of eddies along the magnetic field (Chandran 2000,
 Yan \& Lazarian 2002).

Apart from cosmic rays, acceleration of charged interstellar dust is
also dominated by fast modes (Lazarian \& Yan 2002, Yan \& Lazarian
2003). In stellar magnetospheres with spiral magnetic fields,  
fast-modes possibly play a role in energetics and dynamics 
of winds from usual stars \citep{suz06} and pulsars \citep{lyu03}. 
These issues provide an additional motivation to studies of properties
of fast mode cascade.

The most efficient interaction of fast modes is expected to happen for
waves moving in the same direction (see CL02).  This indicates that,
unlike studies of Alfv\'{e}n modes for which the 3D character of interactions
is essential, 1D simulations could still give, at least in some respects, 
correct physical properties of the fast-mode turbulence.   
1D simulations have a big advantage that we can use a number of grid points 
for a broader inertial range than 2D or 3D simulations.  
Based on these issues, we carry out 1D decay simulations of fast-mode 
turbulences. 
We should cautiously analyze simulation data, because 
the 1D geometry systematically enhances shocks without dilution to the 
tangential directions. 

In what follows we discuss our code in \S 2, calculate power spectra in
\S 3, and study cascading of fast modes in \S 4. We discuss our findings
in \S 5. 

\section{The code and the simulations setup}

We initially give perturbations of fast mode by fluctuations of longitudinal 
velocity, $v_x$, transverse velocity, $v_y$, transverse magnetic field, $B_y$, 
density, $\rho$, and pressure $p$ (or specific energy, $e=\frac{p}{(\gamma-1)
\rho}$, where $\gamma$ is a ratio of specific heat) in magnetically 
dominated plasma, namely
low-$\beta$ conditions.  We dynamically treat time evolution of each
variables by solving ideal MHD equations without continuous driving (decay
simulation).  We do not consider the third ($z$) component of magnetic
field and velocity, which indicates that \Alfven mode is switched off
in our simulation, whereas fast, slow, and entropy modes are automatically
treated. 
We assume one-dimensional approximation, namely all the
physical variables depend only on $x$; a wave number vector has 
only the x-component, namely, $k= k_x$.
The number of grid points of the $x$ component is 65536($=2^{16}$), and 
the periodic boundary condition is imposed in $x$-direction. 

\begin{equation}
\label{eq:mass}
\frac{d\rho}{dt} + \rho\frac{\partial v_x}{\partial x}= 0 , 
\end{equation}
\begin{equation}
\label{eq:mom}
\rho \frac{d v_x}{dt} = -\frac{\partial p}{\partial x}  
- \frac{1}{8\pi}\frac{\partial}{\partial x}  (B_{y}^2)
\end{equation}
\begin{equation}
\label{eq:moc1}
\rho \frac{d}{dt}(v_y) = \frac{B_x}{4 \pi} \frac{\partial B_y} 
{\partial x}.
\end{equation}
\begin{equation}
\label{eq:eng}
\rho \frac{d}{dt}\left(e + \frac{v^2}{2} + \frac{B^2}{8\pi\rho} \right) +  
\frac{\partial}{\partial x}\left[(p + \frac{B^2}{8\pi}) v_x  
- \frac{B_x}{4\pi} (\mbf{B \cdot v})\right]= 0,
\end{equation}
\begin{equation}
\label{eq:ct}
\frac{\partial B_y}{\partial t} = 
\frac{\partial}{\partial x} [(v_y B_x - v_x B_y)], 
\end{equation}
where $B^2 = B_x^2 + B_y^2$ and $v^2 = v_x^2 + v_y^2$; $\frac{d}{dt}$ and 
$\frac{\partial}{\partial t}$ denote Lagrangian and Eulerian time derivatives. 

The minimum wave number corresponds to the box size.
We set-up $p=1$, $\rho=1$, $v_x=0$, $v_y=0$
as background conditions.  
We test two cases of the background magnetic fields, $(B_x,B_y)=(9,3)$  
(quasi-parallel) and $(B_x,B_y)=(3,9)$ (quasi-perpendicular), and 
three cases of a ratio of specific heat, $\gamma=5/3$
(adiabatic), 1.1 (nearly isothermal), and 1 (isothermal).

For the simulations we employed 2nd order nonlinear MHD Godunov-MOCCT 
scheme developed by Sano \& Inutsuka (2007, in preparation) 
(see also Suzuki \& Inutsuka 2005; 2006). 
In this scheme, each cell boundary is treated as discontinuity,
for the time evolution we solve nonlinear Riemann shock tube problems with 
magnetic pressure by the Rankine-Hugoniot relations. Therefore, entropy 
generation, namely heating, is automatically calculated from the shock jump 
condition. A great advantage of our code is that no artificial viscosity 
is required even for strong MHD shocks; numerical diffusion is suppressed to 
the minimum level for adopted numerical resolution.

\section{Time Evolution and Power Spectra}
\label{sec:teps}
\begin{figure}
\figurenum{1} 
\epsscale{1.0} 
\plotone{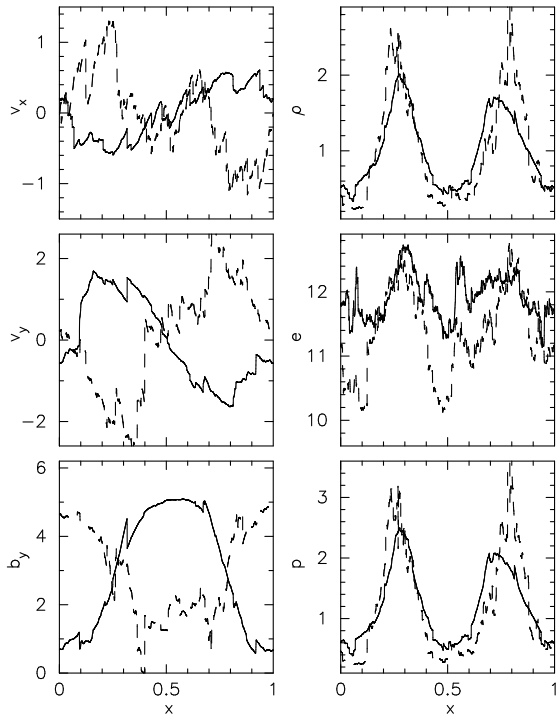}
\caption{Evolution of $v_x$ (top-left), $v_y$ (middle-left), 
$B_y$ (bottom-left), $\rho$ (top-right), $e$ (middle-right), 
and $p$ (bottom-right). 
Dashed and solid lines are the results at $t=0.5$ and $t=3$, respectively. 
The initial spectral slope is $\propto k^{-3/2}$.}
\label{fig:tmev1}
\end{figure}

\begin{figure}
\figurenum{2} 
\epsscale{1.0} 
\plotone{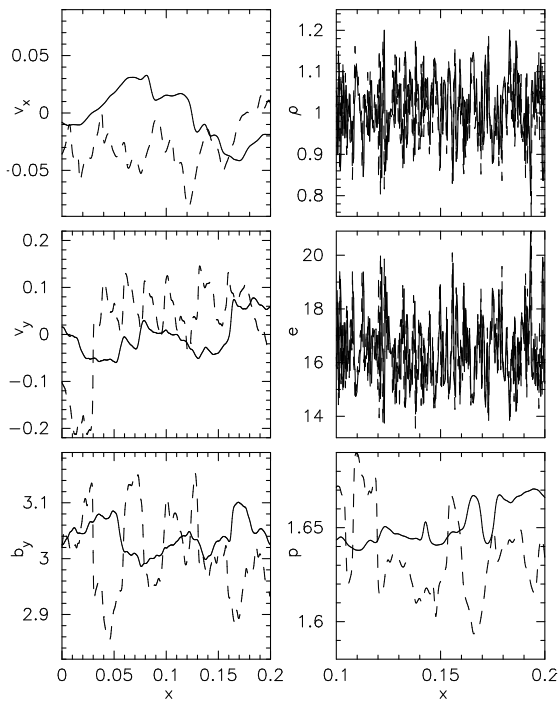}
\caption{The same as Figure \ref{fig:tmev2} but for white noise as the 
initial perturbation. Note that the horizontal axis is enlarged compared with 
Figure \ref{fig:tmev2}; the horizontal axis of the right panels 
is from $x=0.1$ to 0.2, while that of the left panels is from 
$x=0$ to 0.2.}
\label{fig:tmev2}
\end{figure}

We firstly study time-evolution of the simulated turbulences and their 
power spectra. In this section, we present the results of quasi-parallel 
($(B_x,B_y)=(9,3)$) and $\gamma=1.1$ case. 
The initial amplitude is set to be $dB_y=4.4$, corresponding to super-sonic 
and sub-\Alfvenic condition. 
Fast modes traveling in both directions have equal energies, we call this
balanced turbulence. 
We test two types of initial spectral slopes; 
one is $\propto k^{-3/2}$ with 
random $\delta$-correlated phases for all wavenumbers, and the other is 
$\propto k^{0}$, namely white noise. 
Note that $k^{-3/2}$ is a prediction of weak acoustic turbulence, which 
is supposed to have similar propertiles to weak fast-mode turbulence (CL02; 
CL03), whereas the validity of weak turbulence approximation is still 
under debate for acoustic turbulence \citep{fm96,zlf92}.   
Figures \ref{fig:tmev1} \& \ref{fig:tmev2} present time-evolution of 
the cases with initial spectra $\propto k^{-3/2}$ and $k^0$, respectively; 
the six panels 
present $v_x$, $v_y$, $B_y$, $\rho$, $e$, and $p$ at $t=0.5$ (dashed) 
and $t=3.0$(solid).  

\begin{figure}
\figurenum{3} 
\epsscale{1.0} 
\plotone{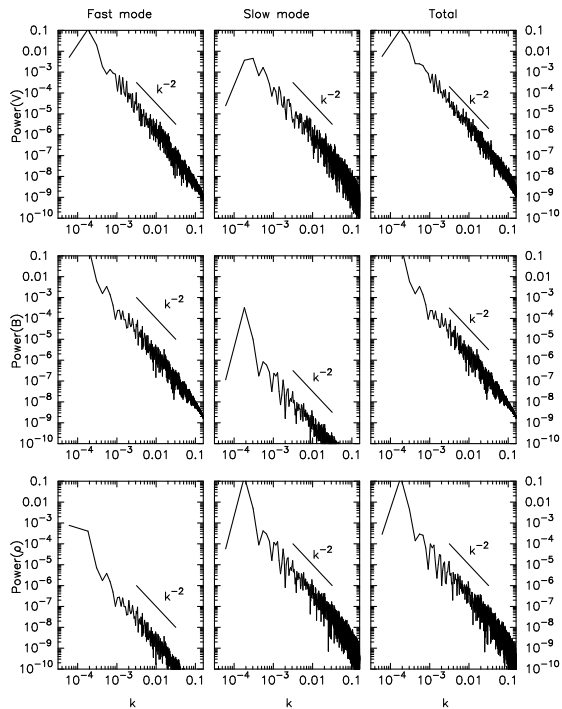}
\caption{
Power spectra of $v$ (top), $B$ (middle), and $\rho$ (bottom) 
decomposed into fast (left) and slow (middle) modes at $t=3.0$ for the intial 
$\propto k^{-3/2}$ perturbation. Composed values are shown on the right 
for comparison. 
}
\label{fig:spev1}
\end{figure}

\begin{figure}
\figurenum{4} 
\epsscale{1.0} 
\plotone{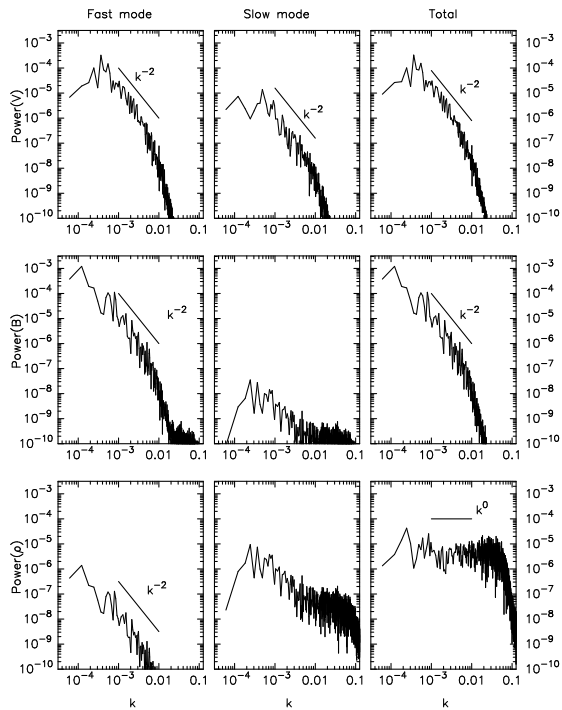}
\caption{Same as Figure \ref{fig:spev1} but for the initial white noise 
perturbation. Nte that the composed density spetrum is mainly owing to 
entropy mode (note show); this is the reason why the total power of 
the density turbulence is larger than the combination of the fast- and 
slow-mode perturbations. The slow mode $B_y$ and $\rho$ spectra might 
contain errors in the high $k$ region (see text).  
}
\label{fig:spev2}
\end{figure}

We also show the power spectra of $v$, $B$, and $\rho$ at $t=3.0$ in 
Figures \ref{fig:spev1} and \ref{fig:spev2}. 
Although our initial conditions contain only fast mode, slow mode 
is generated by nonlinear process of mode interaction. 
In order to understand the interplay of modes, 
we decompose all values into fast and slow perturbations (see CL02). 
This method is statistically correct even for nonlinear waves as discussed in 
CL03.   
In Figures \ref{fig:spev1} and \ref{fig:spev2}, the decomposed spectra of 
fast (left) and slow (middle) modes as well as the raw spectra (right). 
Note that the $B_y$ and $\rho$ spectra of the slow mode in the initial 
white noise case (Figure \ref{fig:spev2}) may contain errors in the small 
scale (large $k$) region. 
As we describe later, the density perturbation is mainly due to entropy-mode 
in this case.  
The slow mode spectrum is calculated from the substraction of the dominated 
entropy mode from the total power, which might leave an error in the slow 
mode. 
The $B_y$ spectrum of the slow mode, which is connected to the slow mode 
density spectrum through the frozen-in condition, might also contain  
non-negligible errors. 

The velocities and magnetic fields exhibit many discontinuities, i.e. 
fast and slow MHD shocks, in both cases 
(the left panels of Figures \ref{fig:tmev1} \& \ref{fig:tmev2}). 
The amplitudes of the white noise case is much smaller 
than those of the $k^{-3/2}$ spectrum case, because the white noise 
perturbations initially contain more energy in smaller scales which 
suffer faster damping. 

Figures \ref{fig:spev1} \& \ref{fig:spev2} indicates that the velocity and 
magnetic field perturbations are carried by fast mode. 
This is first because the slow mode is downconverted from the fast mode, 
and second because the fast wave becomes magnetic mode in low-$\beta$ 
medium. 
The slow wave essentially corresponds to hydrodynamic (acoustic) mode, 
and this is the main reason why the slow-mode perturbation dominates the 
fast-mode in the density spectra, 
in accordance with the earlier claims in 
CL03, Passot \& V\'{a}zquez-Semadeni (2003), and 
Beresnyak et al. (2005), whereas the entropy-mode perturbation is also 
important in the initial white noise case 
as discussed later.  

As expected from many discontinuities in Figures \ref{fig:tmev1} \& 
\ref{fig:tmev2}, $v$ and $B_y$ exhibit shock dominated spectra, 
$\propto k^{-2}$ in both $k^{-3/2}$ (Figure \ref{fig:spev1}) and white noise 
(Figure \ref{fig:spev2}) cases. 
This indicates that the weak cascade solution of acoustic turbulence 
($\propto k^{-3/2}$) 
does not hold for any significant time even though it is provided initially.
The randomness of phases typically assumed in acoustic turbulence or in other
theories of weak turbulence breaks down quickly by nonlinear interaction,
leading to Burgers-like shock-dominated random state.
This is different from the results of the 3D simulation by CL02 in which 
spectrum $k^{-3/2}$ is suggested. 
The difference may be due to the difference of 1D vs. 3D 
cases; 
the effect of shocks tend to become systematically predominant in 1D 
simulations because shocks do not dilute geometrically.  

\begin{figure}
\figurenum{5} 
\epsscale{0.8} 
\plotone{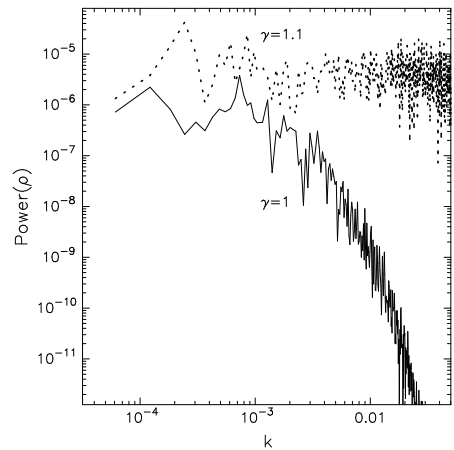}
\caption{Density spectra of the initial white noise perturbations for 
$\gamma=1.1$ (dotted) and 1 (dashed) at $t=3$. }
\label{fig:rhosp}
\end{figure}

Since our simulations do not assume isothermal gas, the different dissipation 
characters influence the evolution of thermal properties. 
The initial specific energy, $e_0(=\frac{p_0}{(\gamma-1)\rho_0})$, is 10 in 
both cases. In the $k^{-3/2}$ case, the heating is localized at first
in places where strong shock heating takes place; at $t=0.5$, $e$ partially 
becomes $\gtrsim 12$, while some places stay $e\approx e_0(=10)$. 
Most places experience the heating eventually and at $t=3.0$ the gas is 
averagely heated up to $e\approx 12$.     
On the other hand, in the white noise case (Figure \ref{fig:tmev2}), 
$e$ is quickly heated to $e\approx 
17$ becauase of the rapid dissipation, and after that $e$ stays 
almost constant with time. 

The densities show different features in Figures \ref{fig:tmev1} and 
\ref{fig:tmev2};    
the $\rho$ structure of the initial $k^{-3/2}$ spectrum case (Figure 
\ref{fig:tmev1}) 
is dominated by larger scale with a couple of shocks, which is
qualitatively simillar to $v$ and $B_y$, while many small-scale structures 
are predominant in the white noise case (Figure \ref{fig:tmev2}). 
Figure \ref{fig:spev1} ($k^{-3/2}$ case) shows the density perturbation is 
mainly carried by the slow-mode and the spectral slope is the shock-dominated 
one, $k^{-2}$, similarly to the $B_y$ and $v$ spectra. 

On the other hand, in the initial white noise case, the total power 
of the density fluctuation (the right bottom panel of Figure \ref{fig:spev2}) 
is much larger than the sum of the fast- and slow-mode perturbations. 
In fact, the density perturbation in this case is mostly owing to an entropy 
mode, 
which is a zero-frequency mode with unperturbed $p$ and with perturbed $\rho$ 
and $e$ (see e.g. Lithwick \& Goldreich 2001).  
By a close look at Figure \ref{fig:tmev2} one can see that the phases of $e$ 
and $\rho$, 
both of which show small-scale structure, are anti-correlated so that $p 
(\propto \rho e)$ shows a smooth feature.
The initial perturbations (white noise) contain sizable energy in the 
small scale. 
These small-scale turbulences quickly dissipate, rather than converted 
to slow-mode perturbation, to entropy-mode fluctuation. 
The density fluctuation of the entropy mode drives the fluctuation 
of the energy deposition by shocks. The heating rate per unit mass is larger 
in hotter regions because of the lower density, which self-sustains the 
entropy-mode perturbation.    
As a result, the density perturbation is preserved with keeping the initial 
flat spectrum until the end of simulation ($t=3$), when the turbulent 
energy of the fast mode decays $\simeq 1/1000$ of the initial value. 

In conclusion, the density perturbation is controlled by initial driving 
turbulence (see Beresnyak 2005 for 3D case). 
This is a kind of bistability.   
When the energy is given in a larger scale, the density perturbation 
is carried by the slow-mode downconverted from the fast-mode. 
The generated slow-mode shows the shock-dominated spectrum. 
On the other hand, when the energy is input in a smaller scale, the density 
perturbation is self-sustained by the entropy-mode fluctuation, which is 
a result of the dissipation of the fast-mode turbulence. 
The density perturbation triggers the fluctuation of the heating, which 
sustains the entropy-mode fluctuation in itself. 
In this case, an initial spectrum is preserved in the density perturbation. 
  



In isothermal ($\gamma=1$) gas, the entropy mode does not exist because 
$\rho$ must follow unperturbed $p$; the $\rho$ spectrum would be different 
especially in a small scale.    
We carry out the simulation of isothermal gas with the initial white noise 
fluctuation.  
Figure \ref{fig:rhosp} compares the density power spectra adopting the 
initial white noise perturbations for $\gamma = 1$(isothermal) and 1.1 
at $t=3$.  
This figure shows that the density spectrum, which is due to the slow-mode 
in the isothermal gas, becomes steep nearly $\propto k^{-2}$.
Therefore, the imprint of the initial spectrum is unique in the density 
perturbations of non-isothermal gas, in which entropy-mode perturbation 
does exist.  
We should note, however, that the fast-mode turbulence is not influenced 
whether we adopt isothermal or nonisothermal conditions. 

The above result shows that density perturbation in a small scale is 
influenced by an equation of state of gas. 
In general, a ratio of specific heat, $\gamma$, is controlled by physical 
processes such as radiative cooling and thermal conduction.   
This shows that we might get some information on gas properties 
from observations of density turbulence. 
Density perturbation is also important in parametric instability because 
it can work as a mirror against Alfv\'{e}n and magnetosonic waves. 
Thus, the relation between $\gamma$ and density turbulence is 
interesting. However, unlike fast-mode turbulence, 3D treatment is 
essential to entropy as well as slow modes, since they are passively 
advected \citep{lg01} and behave very differently in 1D and 3D; 
we pursue this issue in our future project. 
In this paper we concentrate on the evolution of the fast mode  
from now.   

\section{Cascading of Fast Mode}
We analyze time-evolution of integrated energy density (energy column 
density), \\$\int dx(\rho \delta v^2 +\delta B^2)/2$, of the decaying 
fast modes. 

\subsection{Dependence on Initial Amplitude}
\begin{figure}
\figurenum{6} 
\epsscale{0.68} 
\plotone{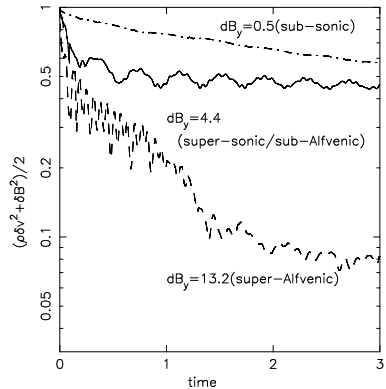}
\caption{Energy of the fast modes of various initial amplitudes. 
The background field strength is $(B_x,B_y)=(9,3)$. 
Dot-dashed, solid, and dashed lines are results of 
$dB_y=0.5$ (sub-sonic), $4.4$ (super-sonic/sub-\Alfvenic), and $13.2$ 
(super-\Alfvenic). The initial energy density is normalized at unity in each 
case. }
\label{fig:dpamp}
\end{figure}

Figure \ref{fig:dpamp} shows dissipation of fast mode energy as a
function of time for different initial amplitudes with the same
background $(B_x,B_y) =(9,3)$. The figure indicates that larger
amplitude gives faster damping. This is reasonable since the
dissipation is owing to nonlinear processes such as shocks and
wave-wave interactions.  Damping times, which are defined as slopes of
the respective lines, become longer at later times as the amplitudes
decrease, which is also consistent with the nonlinear dissipation
through steepening.

\subsection{quasi-Parallel {\it vs.} quasi-Perpendicular}


\begin{figure}
\figurenum{7} 
\epsscale{0.68} 
\plotone{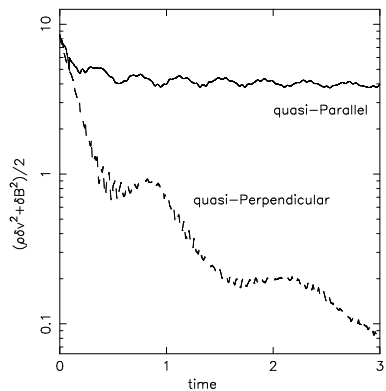}
\caption{Energy (per grid) of the fast modes of the quasi-parallel, 
$(B_x,B_y)=(9,3)$, (solid) and quasi-perpendicular, $(B_x,B_y)=(3,9)$, 
(dashed) cases.}
\label{fig:prpp}
\end{figure}

Figure \ref{fig:prpp} compares the energy of the fast modes of 
quasi-parallel ($B_x=9, B_y=3$) and quasi-perpendicular ($B_x=3, B_y=9$) 
cases. The same $dB_y=4.4$ and $\gamma =1.1$ are adopted. 
The figure shows that the dissipation is more effective in 
the quasi-perpendicular case. This is because the fast mode shows more 
compressive character as an angle between the propagation and the magnetic 
field increases and the shock dissipation is enhanced 
(Suzuki et al. 2006). 

\subsection{Dependence on Plasma-$\beta$}
Since our simulations do not assume isothermal gas, the gas is influenced by 
the heating, and the plasma-$\beta$ values change with time by the energy 
transfer from magnetic fields to gas. This effect is more prominent in 
the adiabatic cases ($\gamma=5/3$).  
We have found from the simulations that the dependence of the decay of the 
fast-mode turbulence on plasma-$\beta$ is very weak provided the plasma 
is kept magnetically dominated ($\beta < 0.5$) during simulations. However, if
$\beta$ approaches unity, the dissipation becomes saturated.  

\begin{figure}
\figurenum{8} 
\epsscale{1.0} 
\plotone{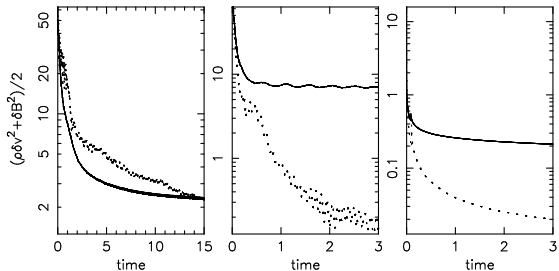}
\caption{{\it left} : 
Energy (per grid) of the fast modes of different $\gamma=1.1$(dotted) and 
5/3 (solid). 
The same background field strength, $(B_x,B_y)=(9,3)$, amplitude, 
$dB_y=13.2$, and the initial spectrum, $\propto k^{-3/2}$, are adopted. 
{\it middle} : the same as the left panel but for $(B_x,B_y)=(3,9)$.
{\it right} : the same as the left panel but for the initial white noise 
spectrum.
}
\label{fig:dbeta}
\end{figure}

Figure \ref{fig:dbeta} compares cases with the same $(B_x,B_y)=(9,3)$(left) 
or $(B_x,B_y)=(3,9)$(right) and
$dB_y=13.2$ but different $\gamma=1.1$ (dotted) and 5/3 (solid).  In
the case with $\gamma=5/3$ the plasma is heated up so that the gas
pressure eventually becomes comparable with the magnetic pressure
($\beta\simeq 1$). In contrast, the plasma $\beta$ stays well below
unity in the cases with $\gamma=1.1$ because the heating is less
efficient due to the smaller $\gamma$.  All the three panels of 
Figure \ref{fig:dbeta} show
that once the plasma-$\beta$ becomes unity the dissipation is suppressed in
the case with $\gamma=5/3$, while the energy density monotonically
decreases in the case with $\gamma=1.1$.
In the middle and right panels, the fast-mode energies of 
the $\gamma=1.1$ cases are far below those of the $\gamma=5/3$ cases. 
Even in the left panel, the energy of the $\gamma=1.1$ case would be smaller 
than that of the $\gamma=5/3$ case, if we proceeded the simulation longer 
time.

At later epochs in the case with $\gamma=5/3$, slow mode also has 
a sizable amount of energy because the phase speed of the slow mode 
becomes comparable with that of the fast mode. 
This tendency is more extreme in the quasi-perpendicular case (the right 
panel), because the plasma is heated up more due to larger compressibility.  

\subsection{Balanced {\it vs.} Imbalanced}

\begin{figure}
\figurenum{9} 
\epsscale{0.68} 
\plotone{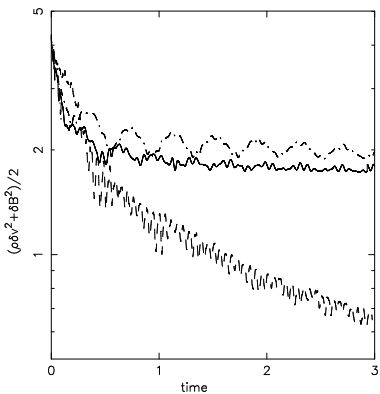}
\caption{Comparison of balanced and imbalanced fast-mode turbulence. 
Dot-dashed line indicates total fast mode energy in balanced case with 
$dB_y=4.4$, solid line indicates fast mode energy propagating to one direction 
in balanced case with $dB_y=6.2$, and dashed line denotes fast mode energy 
in imbalanced case with $dB_y=4.4$. 
}
\label{fig:blib}
\end{figure}

Figure \ref{fig:blib} compares the evolution of fast modes in balanced 
and fully-imbalanced cases. 
The background field strength is $(B_x,B_y)=(9,3)$. 
Dot-dashed line indicates total fast mode energy in balanced case with 
$dB_y=4.4$, solid line indicates fast mode energy propagating to one direction 
in balanced case with $dB_y=6.2$, and dashed line denotes fast mode energy 
in imbalanced case with $dB_y=4.4$.  
Note that the energy is proportional to $dB_y^2$ so that the initial 
values of these three lines are the same.  
The figure shows that 
the imbalanced cascade is more dissipative, which is opposed to 
the tendency obtained in the \Alfvenic turbulences.  

\begin{figure}
\figurenum{10} 
\epsscale{1.0} 
\plotone{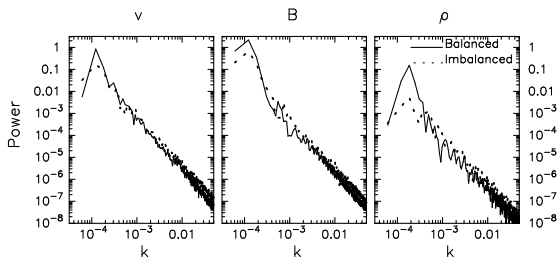}
\caption{Power spectra of the balanced (solid) and balanced (dotted) 
turbulence at $t=3$.}
\label{fig:bibksp}
\end{figure}

One reason is different nature of the dissipation of fast and \Alfven 
modes. 
The fast mode is isotropic. The dissipation occurs 
among fast waves which meet the following resonance conditions (see CL02): 
\begin{equation}
\omega_1 + \omega_2 = \omega_3 ,
\end{equation} 
\begin{equation}
\mbf{k}_1 + \mbf{k}_2 = \mbf{k}_3 .
\end{equation} 
Since, $\omega \propto k$ for the fast modes, the resonance conditions can 
be satisfied only when all three $\mbf{k}$ vectors are collinear.  
When parent '1' and '2' waves propagate in the counter directions, 
(different signs of $\mbf{k}_1$ and $\mbf{k}_2$) 
the daughter wave should have smaller wavenumber, $\mbf{k}_3$, (larger 
wavelength). 

In contrast, if parent waves propagate in the same direction, wavenumber of 
the daughter wave becomes larger than the parents', which is more appropriate 
for the dissipation. In the imbalanced case, we only have fast modes traveling 
in the same direction so that the resonances selectively generate waves with 
shorter wavelengths which suffer faster dissipation. 
Dot-dashed (energy of one direction in balanced case with $dB_y=6.2$) and 
dashed (imbalanced case with $dB_y=4.4$) lines show very similar damping 
at the beginning ($t\le 0.5$). 
This indicates that the dissipation of 1D fast mode 
turbulence is controlled by wave packets traveling in the same direction 
and almost independent of those in the counter directions during this period, 
since the initial energy of one direction in the former case is identical to 
the initial total energy of the latter. 

In later time, however, the dissipation of the balanced cases is slower 
than the dissipation of the imbalanced case. 
To investigate this difference, we compare the power spectra at $t=3$ 
(Figure \ref{fig:bibksp}). 
The comparison shows that more power is kept at very small 
$k\approx 10^{-4}$ in the balanced turbulence, although both show 
the shock-dominated spectra, $\propto k^{-2}$, in the higher $k$ regime.
This might indicate that a kind of inverse cascade takes place preferentially 
in the balanced turbulence. 
Since turbulences with larger scales (smaller $k$) suffer less damping, 
further dissipation seems suppressed in the balanced turbulence.  
However, we think that to discuss this issue the 1D simulation is not 
sufficient, hence, we leave it to our future studies.      

We also suspect that bulk acceleration affects the faster dissipation of the 
imbalanced fast-mode turbulence. 
The imbalanced flux of fast mode transfers the momentum flux to the fluid. 
As a result, the system is accelerated into the direction 
corresponding to the initial wave momentum flux. Generally, energy density of 
waves is lost in accelerating fluid by transfer of the momentum flux to the 
gas even though the waves themselves do not suffer damping \citep{jaq77}. 
However, after comparison of the bulk momentum flux of the gas with the 
dissipated energy of the fast-mode turbulence, we have found that the 
effect of the bulk acceleration is not so large ($<10\%$) in our case. 


\section{Summary and Discussions}

We have shown that the imbalanced fast modes dissipate in a different way than
\Alfven ones. For the \Alfven turbulence cascading is due to wave packets
moving in the opposite direction, therefore, the imbalance makes the
cascading less efficient. For fast modes
waves moving in the same direction interact more efficiently. We
confirmed that the oppositely moving fast modes are not essential for the
cascading as discussed earlier (see CL02).

However, the obtained power spectrum ($k^{-2}$) 
is different from the weak acoustic turbulence ($k^{-3/2}$) suggested in 
CL02, which may be owing to the different geometries (1D vs. 3D) as discussed 
in \S \ref{sec:teps}. We should note that this problem of the scaling 
of acoustic turbulence is not settled down 
because the approximation of weak cascade is controversial 
\citep{fm96,zlf92}.     
It is clear that further research by 3D simulations with a sufficient 
inertial range is necessary.

We have also found that as the amplitude of compressions increase, the
non-linear damping of turbulence speeds up. In our adiabatic calculations
the restoring force was increasing due to medium heating and as the result
the non-linear steepening and cascading slowed down. Similarly, for the 
modes that compress both magnetic field and gas, the rate of dissipation was
less compared to the modes that mostly compress the gas.

The density perturbation is mainly carried by slow mode perturbation 
in the large scale and entropy mode in the small scale perturbation 
in non-isothermal gas. 
The slow-mode density spectrum is dominated by shocks, $\propto k^{-2}$, 
while the entropy-mode, which is a consequence of the dissipation of the 
initial fast-mode turbulence, preserves the initial turbulent spectrum. 
As a result, the spectral index of the density perturbation is subject 
to the initial spectrum of given fast-mode turbulence. 
In isothermal gas, density spectrum is different in a small scale because 
entropy-mode fluctuation does not exist. 
This shows that to study density turbulence one has to carefully treat 
driving properties of turbulence 
and to take into account appropriate thermal 
processes (e.g. Koyama \& Iutsuka 2002; Larson 2005) rather than assume a 
polytropic equation of state. 
While the density turbulence is an interesting issue, 
obviously 3D treatment is required 
since the slow- and entropy-mode perturbation is different in 1D and 3D 
circumstances; we will study this topic in a future paper.        

In this paper we have focused on the fast-mode in low-$\beta$ medium. 
Various interstellar/intergalactic medium is dominated by magnetic pressure 
rather than gas pressure, i.e., low-$\beta$ condition. 
Referring to the tabulation in YL04 (Table 1 in their paper), galactic halo, 
and warm and cold ISM are mildly low-$\beta$ ($\beta=0.1-1$), 
and dark clouds are more magnetically dominated, $\beta=0.01-0.1$. 
Our simulations are directly applicable to these media.  

Magnetospheres of general stars with surface convection (like the sun) 
and compact objects (white dwarfs and neutron stars) are also in 
low-$\beta$ condition, because the density rapidly decreases upwards. 
Moreover, the turbulence tends to be imbalanced there because 
it is mainly generated from central stars and dominated by 
the outward component.  
Our results show that the fast mode is likely to be the culprit in the
heating and acceleration of winds because it dissipates effectively 
even without the inward component. The advantage of the fast-mode
versus Alfv\'{e}nic turbulence in the stellar wind acceleration
is that the latter requires production of counter-waves via reflection
or by some other mechanism.    

Cascading of imbalanced fast modes is important for the problem of streaming
instability evolution, which is the part of the picture of both galactic
leaky box model and the acceleration of particles in shocks.
This instability arises from the 
 flow of particles along the magnetic field direction and reflects the 
particles back (Kulsrud \& Pearce 1968). Our results
indicate that the fast mode cascading limits the instability even in the
absence of both damping and the ambient turbulence (YL02,
Farmer \& Goldreich 2004). 

In our simulation the dissipation is mainly due to MHD shocks besides 
sub-grid scale damping. However, if we discuss the dissipation in more detail, 
we need to take into account the microphysics. Galactic halo and stellar 
winds generally consist of collisionless plasma. In such conditions, the fast-mode 
as well as slow-mode fluctuations possibly dissipate by transit-time damping 
(e.g. Barnes 1966; Suzuki et al.2006). 
In warm ionized ISM, the dissipation is mainly due to Coulomb collisions. 
If there is a considerable fraction of neutral atoms, e.g. in cold ISM
and dark clouds, collisions between neutrals and ions dominate (see Table 1 in YL04). 

Turbulence may be very different in 1D and 3D. We find, however, that our
attempts to get insight into the dynamics of fast modes using 1D model is
meaningful, as the possible transversal deviation $\delta k$ is limited.
Assuming that the fast modes follow the acoustic cascade with
the cascading time
\begin{equation}
\tau_{k}=\omega/k^{2}v_{k}^{2}=(k/L)^{-1/2}\times V_{ph}/V^{2},
\label{fdecay}
\end{equation}
YL04 used the 
uncertainty condition $\delta \omega t_{cas}\sim 1$,
where $\delta \omega\sim V_{ph} \delta k (\delta k/k)$
to estimate $\delta k$
\begin{equation}
\delta k/k\simeq \frac{1}{(kL)^{1/4}}\left(\frac{V}{V_{ph}}\right)^{1/2}.
\label{dk}
\end{equation}
Eq. (\ref{dk}) provides a very rough estimate of how much the randomization
of vectors is expected due to the fast mode cascading. This estimates suggest
that deep down the cascade where $kL\gg 1$ the approximation of waves moving
in the same direction could be accurate enough.

In this paper we dealt with fast modes irrespectively of the evolution
of the other modes. This possibility corresponds to the earlier theoretical
and numerical studies (GS95, Lithwick \& Goldreich 2001, CL02, CL03). 
Recently, however, 
Chandran (2005) discussed the interaction between fast and 
Alfv\'{e}n modes under the approximation that the latter constitute
 {\it weak} turbulence, i.e. turbulence that cascades slowly due to
weak non-linearities. 
We expect appreciable mitigation of the Alfv\'en-fast mode interaction
compared to those in Chandran (2005), when Alfv\'{e}nic 
turbulence is {\it strong}, i.e. when the turbulence evolves fast, over
just one wave period. We feel that the resulting large disparity in the rate of
evolution of the Alfv\'enic and fast modes should result in much less cascading
of fast modes by the Alfv\'enic modes compared to the case of the weak
Alfv\'enic turbulence. Note, that
 strong Alfv\'enic turbulence is default for most
astrophysical situations, while the inertial range of the weak Alfv\'enic
turbulence is limited\footnote{It is easy to see that if the injection happens at scale $L$
with Alfv\'en Mach number $M_A=V/V_A$, where $V_A$ is the Alfv\'en velocity
that arises from the total magnetic field in the fluid, the range for weak
turbulence is limited by $[L, LM_A^2]$. While the regime of weak Alfv\'enic
turbulence can still be important
(see Lazarian 2006), in most astrophysical situations that regime of
strong  Alfv\'enic turbulence covers much longer inertial range.} 
(Goldreich  \& Sridhar 1997, Lazarian \& Vishniac 1999). 
In our next paper we shall provide calculations of the fast modes interaction 
with the Alfv\'enic modes.  

{\bf Acknowledgments}

T.K.S. is supported by JSPS Research Fellow for Young Scientist, 4607, and 
a Grant-in-Aid for Scientific Research, 18840009, from MEXT of Japan. 
AL acknowledges the support of the NSF Center for 
Magnetic Self-Organization in Laboratory and Astrophysical Plasmas and the NSF 
grant AST-0307869 and NSF-ATM -0312282. 
AB acknowledges the support of the Ice Cube project.

\end{document}